\documentclass{aastex631}

\begin{document}


%\title{The carbon star DY Persei is a very cool R Coronae Borealis variable}
\title{The carbon star DY Persei may be a cool R Coronae Borealis variable} 
%% Use \author, \affil, plus the \and command to format author and affiliation 
%% information.  If done correctly the peer review system will be able to
%% automatically put the author and affiliation information from the manuscript
%% and save the corresponding author the trouble of entering it by hand.
%%
%% The \affil should be used to document primary affiliations and the
%% \altaffil should be used for secondary affiliations, titles, or email.

%% Authors with the same affiliation can be grouped in a single

%\author{P. Ventura et al.}

%\author{D. A. Garc\'{\i}a--Hern\'andez\altaffilmark{1,2}}

%\author{N. Kameswara Rao\altaffilmark{3}}

%\author{David L. Lambert\altaffilmark{4}}

%\author{K. Eriksson\altaffilmark{5}}

%\and

%\author{A. B. S. Reddy\altaffilmark{3}}

%\altaffiltext{1}{Instituto de Astrof\'{\i}sica de Canarias (IAC), E-38205 La Laguna, Tenerife, Spain}
%\altaffiltext{2}{Departamento de Astrof\'{\i}sica, Universidad de La Laguna (ULL), E-38206 La Laguna, Spain}
%\altaffiltext{3}{Indian Institute of Astrophysics, Bangalore 560034, India}
%\altaffiltext{4}{The W. J. McDonald Observatory \&  Department of Astronomy, University of Texas at Austin, TX 78712, USA}
%\altaffiltext{5}{Theoretical Astrophysics, Department of Physics and Astronomy, Uppsala University, Box 516, 751 20, Uppsala, Sweden}

\correspondingauthor{D. A. Garc\'{\i}a-Hern\'andez}
\email{agarcia@iac.es}

\author[0000-0002-1693-2721]{D. A. Garc\'{\i}a-Hern\'andez}
\affiliation{Instituto de Astrof\'{\i}sica de Canarias, C/ Via L\'actea s/n, E-38205 La Laguna, Spain}
\affiliation{Departamento de Astrof\'{\i}sica, Universidad de La Laguna (ULL), E-38206 La Laguna, Spain}

\author{N. Kameswara Rao}
\affiliation{Indian Institute of Astrophysics, Bangalore 560034, India}

\author{David L. Lambert}
\affiliation{The W. J. McDonald Observatory \&  Department of Astronomy, University of Texas at Austin, TX 78712, USA}

\author{K. Eriksson}
\affiliation{Theoretical Astrophysics, Department of Physics and Astronomy, Uppsala University, Box 516, 751 20, Uppsala, Sweden}

\author{A. B. S. Reddy}
\affiliation{Indian Institute of Astrophysics, Bangalore 560034, India}

\author{Thomas Masseron}
\affiliation{Instituto de Astrof\'{\i}sica de Canarias, C/ Via L\'actea s/n, E-38205 La Laguna, Spain}
\affiliation{Departamento de Astrof\'{\i}sica, Universidad de La Laguna (ULL), E-38206 La Laguna, Spain}

\begin{abstract}

Optical and near-IR photometry suggests that the carbon star DY Persei exhibits fadings similar to those of R Coronae Borealis (RCB) variables. Photometric surveys of the Galaxy and Magellanic Clouds uncovered new DY Per variables with infrared photometry identifying them with cool carbon stars, perhaps, with an unusual tendency to shed  mass. In an attempt to resolve DY Per's identity crisis -- a cool carbon giant or a cool RCB variable? --   we analyze a high-resolution H\&K band spectrum of DY Per. The CO first-overtone bands in the K-band  of DY Per show a high abundance of $^{18}$O such that $^{16}$O/$^{18}$O $= 4\pm1$, a ratio sharply at odds with published  results for `regular' cool carbon giants with $^{16}$O/$^{18}$O $\sim 1000$ but this exceptionally low ratio is  characteristic of RCB-variables and HdC stars.  This similarity  suggests that DY Per indeed may be a cool RCB variable. Current opinion considers  RCB-variables to  result from  merger of a He onto a CO white dwarf; observed abundances of these H-deficient stars including the exceptionally  low $^{16}$O/$^{18}$O ratios  are in fair accord with predicted compositions for white dwarf merger products.  A H-deficiency for DY Per is not directly observable but is suggested from the strength of a HF line and an assumption that F may be overabundant, as observed and predicted for RCB stars.

\end{abstract}




%\section{Introduction} \label{sec:intro}
\section{Introduction} 

DY Per, a  cool carbon N-type  star, was suggested by Alksnis (1994) to be a  R CrB
variable (RCB) on account of its  RCB-like fadings at unpredictable times, an extreme 
behavior unusual among the common N-type carbon stars.   As a result of  photometric
surveys of regions of the Galaxy and the Magellanic Clouds, new  identifications of RCB
and DY Per  variables have been proposed.  An initial distinction between RCB and DY Per
variables in photometric surveys  appears to be based on the observation that these
variables exhibit  rather similar declines and recoveries  but RCB variables show the
faster declines and slower recoveries (Alcock et al. 2001).  The distinction between RCB
and DY Per variables was amplified when infrared photometry was undertaken. For example,
the two color plot J-H vs H-K (or  an equivalent)  shows DY Per variables at maximum
light are located among the common N-type carbon stars and well separated from RCB
variables  at their maximum light- see, for example,  Alcock et al. (2001, Fig. 9),
Soszy\'{n}ski et al. (2009, Figs. 1 \& 3),  and Tisserand et al. (2009, Fig. 4). In
addition,  variables in  the Magellanic Clouds  clearly show that  in the 
color-magnitude  $M_V$ - $(V-I)$ diagram the DY Per variables are well mixed in with the
N-type carbon  stars at about $M_V \sim -2$ and $(V-I) \sim 2.2$ but the RCB variables
run from $M_V \sim -2$ at $(V-I) \sim 1.5$ to $M_V \sim -5$ at $(V-I) \sim 0.0$.  This 
sequence encourages the view that DY Per variables and the RCB variables may form a run
of connected  variables at a  similar luminosity but covering a range of effective
temperatures\footnote{Note that, very recently, Crawford et al. (2023) establishes
a spectral classification system for the RCB and HdC stars and discusses how DY Per
might fit into this system as well as its possible temperature.}. Photometric   and
low-resolution optical spectroscopic correspondence between DY Per variables and common
N-type carbon stars have also encouraged  ideas  that DY Per variables are  unusual
variants of  normal carbon stars. For example, Soszy\'{n}ski et al. (2009)  declare
`our candidate DY Per stars form a continuity with other carbon-rich long-period
variables, so it seems that DY Per stars do not constitute a separate group of variable
stars'.  And Tisserand et al.  (2009) consider that `[DY Pers] are ordinary carbon stars
with ejection events. However, more spectroscopic observations and abundance analysis
will be necessary to really answer this question.'  This question is the principal focus
of our analysis of a high-resolution infrared spectrum.

In this paper, we discuss a high-resolution infrared spectrum of DY Per across the K band acquired with the novel spectrograph IGRINS  (Park et al. 2014).   Key spectroscopic indices  $^{16}$O/$^{18}$O and $^{16}$O/$^{17}$O  from the CO first-overtone vibration-rotation bands  measurable from our spectra   provide the opportunity to test the degree to which DY Per is related either to ordinary carbon stars or to the RCB variables.  Published analyses of CO lines in high-resolution  K band spectra of ordinary carbon stars show $^{16}$O/$^{17}$O $\sim$ $^{16}$O/$^{18}$O $\sim 1000$ (see below for details) as predicted theoretically for C-rich low mass AGB stars. In sharp contrast, cool RCB variables and  those HdC stars with CO bands in their K band spectra show remarkable enhancements of $^{18}$O beginning with   Clayton et al. (2005)  reporting  a startling $^{16}$O/$^{18}$O $ \leq 1$  for the HdC HD 137613. Such  selective enhancement of $^{18}$O but not $^{17}$O is taken to be a key to identification of DY Per with the family of   HdC and RCB stars, all warmer than DY Per.   A low   $^{16}$O/$^{18}$O  isotopic ratio   has generally been taken as a signature  of the  merger of a He white dwarf with a C-O white dwarf.  Such a   low   $^{16}$O/$^{18}$O  ratio is exceptional  among stellar atmospheres  and occurrence of DY Per's low ratio  among low ratios for RCB variables and HdC  stars appears to be  an indicator of DY Per's link to these H-deficient stars.  We search our IGRINS spectrum of DY Per for additional products of white dwarf mergers including the F abundance obtainable  from HF vibration-rotation lines in the K band but our search is thwarted by the inability to obtain  independent estimates of the  star's H  and F abundances.  

\section{IGRINS  spectra}

Our high-resolution H\&K band spectrum of DY Per was obtained on 2014 December 4 with the 107-inch Harlan J. Smith telescope at The W.J. McDonald Observatory using the IGRINS spectrograph (Park et al. 2014).   Spectra of relevant stars were also obtained in  2014 December or in a 2016 June-July observing run.  Observations included  the  HdC star HD 137613,  classical  RCBs  -- XX Cam, UV Cas, R CrB and SU Tau -- observed at or near maximum light,  and the M giant $\beta$ Peg.  None of the observed RCBs   show photospheric CO lines  and, hence, are not of primary interest to our investigations of the oxygen isotopic ratios.  

An IGRINS spectrum covers the whole H (1.49$-$1.80 $\mu$m) and K (1.96$-$2.46 $\mu$m)
bands simultaneously  with a (2kx2k) HAWAII-2RG array as detector for each camera. 
Basic data reduction (i.e., sky subtraction, flat-fielding, bad pixel correction,
aperture extraction, and wavelength calibration) were performed with the IGRINS
Pipeline Package. Telluric features were removed with the help of the spectrum of a
hot star observed at a very similar airmass immediately following observation of
the program star: removal of telluric lines from the stellar spectrum is generally
 complete except for  the strongest telluric lines. The spectral
resolving power as estimated from the Gaussian full width at half maximum of weak
telluric lines at 22200 \AA\ in the comparison star is about 45,000, as designed
and built. Standard air wavelengths are given here throughout. Stellar lines have
been corrected for the radial velocity. Fortunately,  these luminous variable stars provide absorption lines with symmetrical profiles without either a hint of  splitting or a velocity variation with line strength, as may occur in some luminous stars such as Mira variables, for example.

Prior to the IGRINS observing runs, we had obtained and analyzed high-resolution spectra in selected portions of the K band of  five HdC and nine RCB stars (Garc\'{i}a-Hern\'{a}ndez et al. 2009, 2010). Several HdC and RCB stars  showed first-overtone CO vibration-rotation bands and yielded  measures of the $^{16}$O/$^{18}$O ratio as well as providing measurements of  lines from the CN Red, C$_2$ Phillips and Ballik-Ramsay systems, all three  systems  contaminate  the CO first-overtone bands.  These spectra from the PHOENIX spectrograph at Gemini South (Hinkle et al. 2003)  are at a   resolution ($R = 50000$)  and  a continuum S/N ratio similar  to our IGRINS spectra. Selective use of the PHOENIX spectra will be entertained here. 

Finally, the spectrum of the N-type carbon star TX Psc was taken from  the collection of high-resolution H\&K-band spectra of carbon stars obtained with the Kitt Peak FTS (Lambert, Gustafsson, Eriksson, Hinkle  1986), also   at a similar spectral resolution and S/N ratio to the IGRINS spectrum of DY Per.  Various results  gleaned from this collection of infrared spectra will be referred to below. 

Two portions of the K band   (Fig.  1)  serve to illustrate the quality of our IGRINS  spectra  but also highlight a key difference  between DY Per and  TX Psc, a typical carbon giant.  In Fig.  1's left-hand  panel, the spectrum of both DY Per and TX Psc is provided by primarily a mix of CN  and C$_2$ lines. CN lines are of similar strengths in the two stars.  C$_2$ lines are so much stronger in DY Per  that outstanding examples  may be easily  distinguished.   But the vital point is that the strongest CN or C$_2$  lines  in DY Per and TX Psc have about the same depth, i.e., $I  \sim 0.5$ where the  intensity  $I$ is  referenced to unity at the local continuum.   In Fig.  1's right-hand panel,   DY Per and TX Psc  are compared in  a region dominated by  CO lines  exclusively from  the 2-0 $^{12}$C$^{16}$O  band; all other bands whatever the  CO isotopologues contribute  at longer wavelengths. The CO contribution is    contaminated   with  C$_2$, and CN  lines which, as anticipated, are  intrinsically  weaker at  these longer wavelengths.  Especially striking in this panel is the greater depth of the CO lines in TX Psc than in DY Per: $I \simeq 0.1$  for TX Psc but $I \sim  0.4$ for DY Per.   Constant  but different $I$  for TX Psc and DY Per are  maintained for the strongest $^{12}$C$^{16}$O lines across the K band. For DY Per, the minimum $I$  for the strongest  lines of C$_2$,  CN  (Fig.  1's left-hand panel)   and CO (Fig. 1's right-hand panel )  are  similar.    DY Per is here set apart from TX Psc and similar carbon stars.  This difference in minimum depths of strong CO lines is  taken to reflect primarily a lower O abundance for DY Per than TX Psc.

%One  wishes to understand how the contrasting minimum $I$ of CO lines in DY Per and TX Psc  result from their differences in atmospheric structure or chemical composition. 

%1
%Figure 1 with the observed K-band spectra
%Please change x-axis to microns from 10^4Angstroms
\begin{figure*}
\includegraphics[angle=-90,scale=.3]{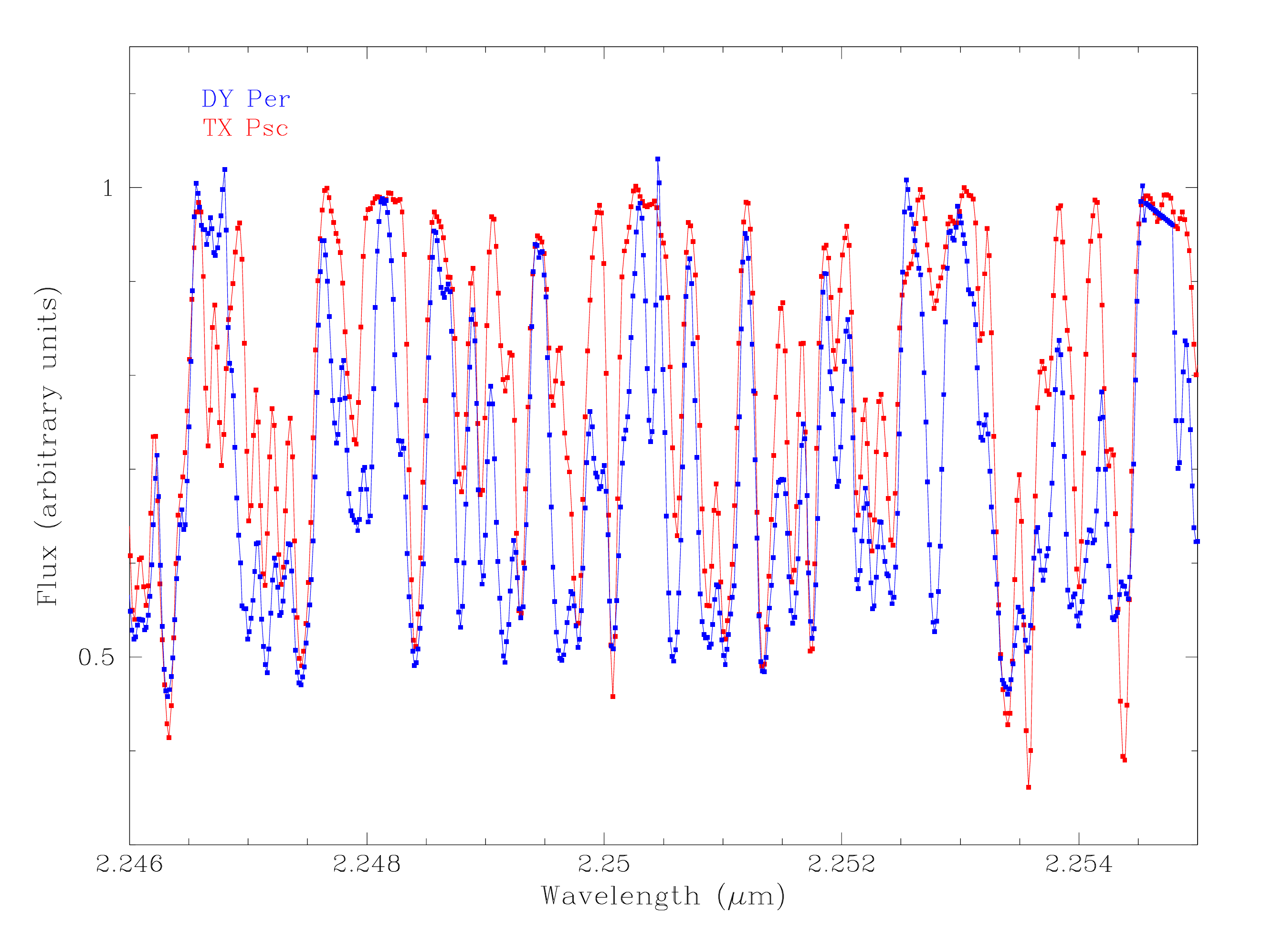}%
\includegraphics[angle=-90,scale=.3]{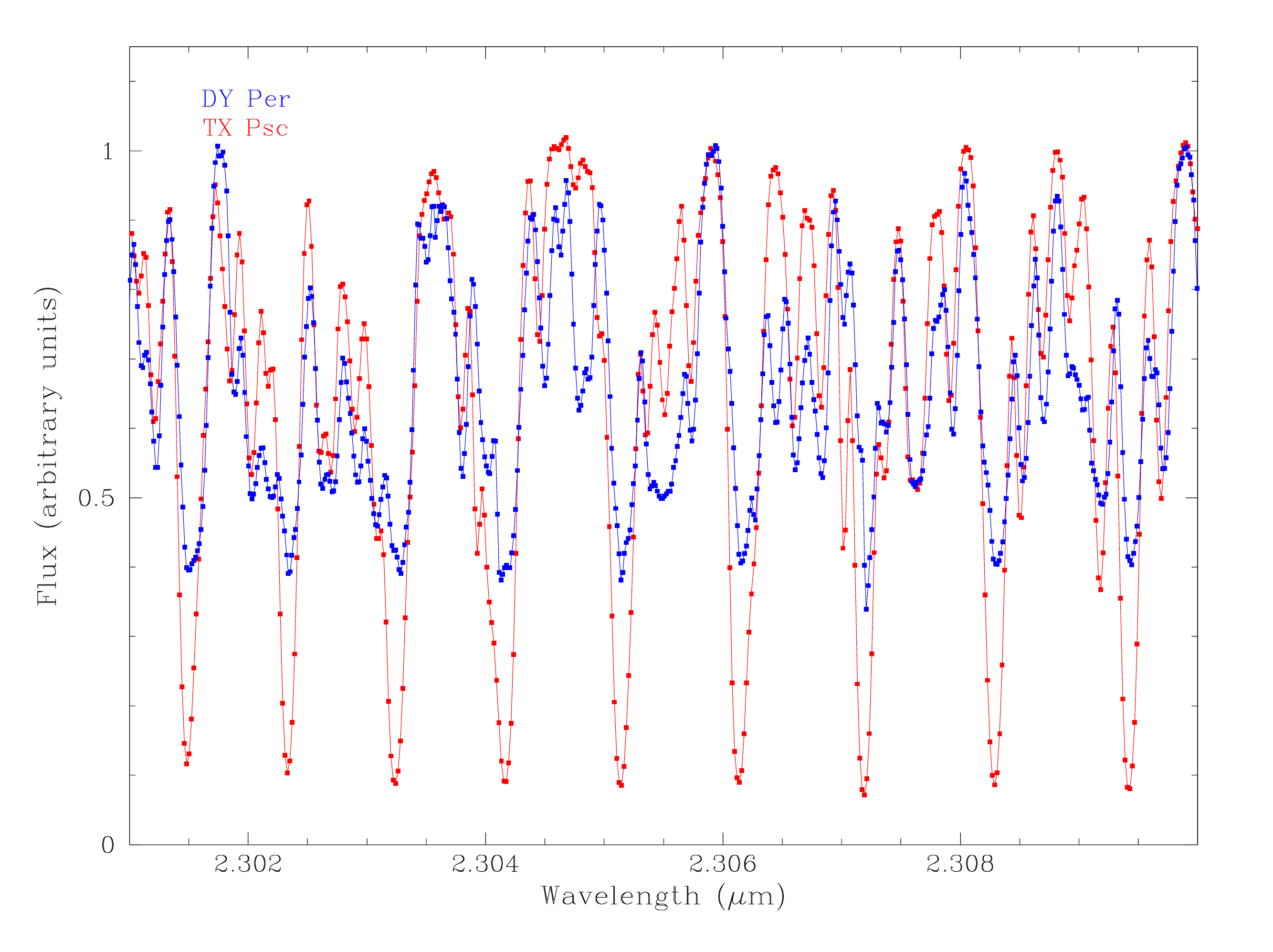}
\caption{Left-hand panel: Spectrum of DY Per (in blue) and TX Psc (in red) from 2.246$\mu$m to 2.255$\mu$m showing a mix of CN and C$_2$ lines in a region free of CO lines.  Inspection shows immediately that  DY Per's C$_2$ lines are much stronger than TX Psc's. Right-hand panel: Spectrum of DY Per (in blue) and TX Psc (in red) from 2.298$\mu$m to 2.310$\mu$m dominated by 2-0 R branch $^{12}$C$^{16}$O lines which are much stronger in TX Psc than in DY Per. \label{fig1}}
\end{figure*}

%\section{Chemical clues to DY Per's  identity?}

\section{Key analytical tools}

Our  goal  is to determine if DY Per is linked more closely to a carbon AGB giant star or to the family of  H-deficient RCB  and HdC stars or may be  the prototype of a new class of rare C-rich giant variable.  Abundance analysis of a cool carbon-rich giant is a daunting prospect whatever the chosen wavelength intervals;  many abundance goals routinely attainable for warmer stars  must be set aside or compromised  when analyzing cool carbon-rich giants.   Our particular hope for   IGRINS spectroscopy was directed at   determinations of the  $^{16}$O/$^{18}$O ratio  and  the  hydrogen-deficiency.  Both quantities  afford an  opportunity to suggest the identity of  DY Per's closest relative.   Our  spectroscopic tests  are  supported by  synthetic spectra computed from  model atmospheres and a detailed line list. Here, we describe the adopted model atmospheres, calculation of synthetic spectra, and selection of stellar parameters for DY Per.    

\subsection{Atmospheric models and synthetic spectra}

Model atmospheres adopted here are based on MARCS models introduced by Asplund et al. (1997). The model family was extended for the analysis of carbon AGB stars discussed by Lambert et al. (1986). Modifications were made to the models for analysis of warm RCB stars by Asplund et al. (2000) with in particular adaptation of the models to the stars' severe H-deficiencies. Such models were also used for  analyses of cool RCB and HdC stars discussed by Garc\'{\i}a-Hern\'{a}ndez et al. (2009, 2010) which focused on the $^{16}$O/$^{18}$O ratio. A few additional models were computed for use with DY Per. Quantitative estimates of our principal clues to DY Per's identity are not critically dependent on our choice of model atmosphere.

A collection of spherically-symmetric MARCS model atmospheres (Gustafsson et al.
2008) was  expanded with  atmospheres in the temperature ($T_{\rm eff}$)  range of
3000 to 4000 K and surface gravities (g) in the  $\log$ g range of 0.5 to 1.5 with a
solar-like mix of metal elements. The metallicity of DY Per was described as solar
and lacking $s$-process enrichment by Za\v{c}s et al. (2005, 2007) from  visual
comparisons of high-resolution optical spectra of DY Per near-maximum light with
spectra of  carbon giants. Sample H-deficient atmospheres are included in our grid. 
%Note that adopted H abundances are several orders of magnitude greater than
%indicated for a typical warm RCB like the eponym R CrB with a H abundance $\sim$7.0.
%Based on detection of a key HF line, we find below that DY Per may not be seriously
%H-deficient. 
The models representing DY Per and other carbon stars include opacities
unimportant for warm RCB variables, e.g., contributions from polyatomic molecules
like HCN and C$_2$H$_2$. 

For the computation of synthetic spectra, we adapted the procedure used  in our
studies of the high-resolution K-band spectra of warm RCB and HdC stars (see
Garc\'{\i}a-Hern\'{a}ndez et al. 2009, 2010 for more details). The TURBOSPECTRUM
package (Alvarez \& Plez 1998; Plez 2012) is used. Line lists for key molecules were
taken from  the following sources: the CO vibration-rotation transitions from the
HITRAN 2016 database (Gordon et al. 2017), the CN Red system from Sneden et al.
(2014), and the C$_{\rm 2}$ molecule from Yurchenko et al. (2018). For CO, the
isotopologues $^{12}$C$^{16}$O, $^{13}$C$^{16}$O, $^{12}$C$^{17}$O and 
$^{12}$C$^{18}$O are in the line list. For CN and C$_2$, $^{12}$C and $^{13}$C are
both represented. The HF line list from J\"{o}nsson et al. (2014) was used. 

%Column densities of abundant molecules in N-type carbon giants and similar stars are likely greater than those examined in the laboratory even at elevated temperatures and, thus, stellar spectra are likely to provide some lines not represented in synthetic spectra. 

\subsection{Atmospheric parameters}

Parameters include the effective temperature (T$_{\rm eff}$), the surface gravity ($\log g$), the composition including the H, He, C, N, O, F and overall metallicity, and the macro- and micro-turbulent velocities.  

%Our  surprising product -- the  low $^{16}$O/$^{18}$O ratio is  almost independent of the adopted atmospheric parameters.

DY Per's T$_{\rm eff}$ is estimated from its intrinsic infrared colors and a calibration based on cool carbon giants. The star is reddened by interstellar dust along the line of sight but perhaps also by circumstellar dust which may contribute to the star's infrared flux.  Maps of interstellar reddening versus distance in DY Per's direction (Neckel 1967) provided Yakovina, Pugach \& Pavlenko  (2009) with their estimate $A_{\sc V} = 1.4$ mag for an  assumed distance 1.5 kpc.  The recent GAIA parallax confirms the distance (GAIA Collaboration 2021). Inspection of  interstellar reddening maps by Fitzgerald (1968) and Neckel \& Klare (1980) encourages a view that $A_V$ may be larger than  above. Fitzgerald presents maps of reddening $E(B-V)$ by galactic latitude. For the two zones spanning DY Per's latitude, the average $E(B-V) \simeq 0.7\pm0.1$ at 1.5 kpc  with the bulk of this reddening occurring at distances of less than one kpc and appearing constant beyond that distance. Such an estimate for $E(B-V)$ corresponds to $A_{\sc V} \simeq 2.1\pm0.3$. Neckel \& Klare provide a rather more detailed map  divided into oddly shaped areas. The area including DY Per shows that $A_{\sc V} \simeq 1.2$ at 1.5 kpc. Three areas closer to the Galactic plane at about DY Per's latitude give $A_{\sc V} \simeq  2.0$.  These (and neighboring) maps suggest that interstellar reddening of distant ($d \geq 1$ kpc) stars is latitude-dependent. For DY Per,  $A_{\sc V} = 1.5\pm0.2$ is suggested as the interstellar extinction. Tisserand (2012) estimates  $A_{\sc V} = 1.71$ using Schlegel, Finkbeiner \& Davis's (1998) maps of interstellar extinction.  We adopt $A_{\sc V} = 1.6$. 
 
Extensive studies have been made of the effective temperature of carbon giants.  We consider  here that a T$_{\rm eff}$ -- color calibration developed for carbon giants is applicable to DY Per. We adopt a calibration by Bergeat, Knapik \& Rutily (2001), which links intrinsic infrared color indices to temperatures estimated from angular diameter determinations and photometry. In particular, we adopt the calibration based on the intrinsic (J-K) color index. 

%Their temperature calibration is similar to several previous proposals. For example, Bergeat et al.'s temperature recommendations in their Table 10 are generally within $\pm150$K of values recommended by Lambert et al. (1986) on similar principles for their  abundance analyses of IR spectra of 30 carbon giants. In particular, we adopt the calibration based on the intrinsic (J-K) color index.
 
A long-period variable like DY Per changes infrared colors as it varies. Alksnis et al. (2002) report extensive JHKLM measurements as the J magnitude varied from J = 5.9 to 7.1. Alksnis et al. (2009) list more limited measurements for phases at which J faded to 9.0. Not unexpectedly, indices such as (J-H), (H-K) and (J-K) redden as the star fades. In order to determine colors at maximum light, we identify a tight `blue' cluster of 16 observations providing means of J = 5.94 and  K = 4.07 with  (J-H) = 1.13, (H-K) = 0.74 and (J-K) = 1.87. After correction for $A_{\sc V} = 1.6$, the stellar magnitude becomes K = 3.89 and the color (J-K) = 1.60 -- see differential interstellar reddening corrections by Cardelli, Clayton \& Mathis (1989). Bergeat, Knapik \& Rutily's calibration gives T$_{\rm eff} \simeq 2750$ K, an estimate compatible with crude estimates drawn from fits to DY Per's spectral energy distributions over narrow or broad wavelength intervals. Of course, this estimate may be suspect if DY Per is not a typical carbon giant. Presence of a  circumstellar dust shell may also affect this temperature estimate which is presumably a lower estimate. 

%(The 2MASS measurements listed by SIMBAD are similar: J = 6.369 with (J-H) = 1.185, (H-K) = 0.778 and (J-K) = 1.963.) 

Surface gravity of a cool carbon giant may be set from an estimate of stellar mass and a determination of the de-reddened bolometric magnitude or absolute luminosity.  For the GAIA parallax, the bolometric correction $BC_K$ = 3.09 for a carbon star with  (J-K) = 1.60 (Kerschbaum, Lebzelter, Mekul 2010) which with K = 3.89 converts to $M_{bol}$ = -3.9  or $\log L/L_{\odot} =  3.5$. The inferred surface gravity $\log g = -0.3$ where we adopt a solar mass for the star (see below).

DY Per's overall metallicity was estimated by Za\v{c}s et al. (2007) from inspection of high-resolution optical spectra of DY Per and carbon giants of similarly low temperature: (UV Cam, R Lep, W CMa and U Hya) with previously estimated metallicities. Za\v{c}s et al. (2007) gave DY Per's metallicity as solar and the ratio [$s$/Fe] also as solar indicating a lack of $s$-process enrichment. For carbon giants, Lambert et al. (1986) noted the difficulty in identifying atomic lines in  H+K band spectra and provided a brief list of possible `clean' lines in the K band short ward  of the region dominated by CO lines. Consideration of the 1986 line list and examination of the spectra of DY Per and TX Psc along with fitted synthetic spectra of CN and C$_2$ confirm the great difficulty of isolating atomic lines in our K band spectrum.

\subsection{The circumstellar shell, stellar photometry and infrared spectroscopy.}

Our initial assessment of the impact of DY Per's circumstellar dust on its infrared spectrum was provided from the spectral energy distribution (SED) set by optical photometry (B, J, H, K) to infrared  photometry ([3.4], [4.6], [12], [22] -- WISE, [9.0], [18] -- AKARI) for `maximum' light by Tisserand (2012, Fig. 9). This SED and others assembled by Tisserand for RCB and HdC stars were fitted by him using the program DUSTY (Nenkova, Ivez\'{i}c, Elitzur  2000) with a black body to represent stellar emission and with a shell composed of amorphous carbon grains with a size distribution suggested by Mathis, Rumpl \& Nordsieck (1977), at a uniform shell temperature and a visual optical depth $\tau_V$. Obviously, the black body temperature obtained for the star DY Per is no more than a rough approximation to the stellar flux crossed by strong molecular bands. Modeling of the shell emission provides a fair estimate of that shell's temperature and visual optical depth. For DY Per, Tisserand gives $T_{\rm eff} \simeq 3000$K for the star and $T = 1200$ K with  $\tau_{\sc V}$ = 0.36 for the circumstellar shell. Inspection of Tisserand's SED for DY Per shows that the shell flux may be from 0 -- 40$\%$ of the stellar flux at K as set by the 3000 K black body. 
%Thus, even at maximum light the stellar spectrum in the K band may be diluted by emission from the shell. 
For this study, we assume that the shell does not dilute the stellar spectrum at the K band.
% Consideration must be given to the possibility that our IGRINS spectrum was obtained when its H and K magnitudes were below their maximum and our spectrum was more diluted by emission from the shell. This emission is presumed to contribute a continuous spectrum. 

In addition, the circumstellar dust will likely redden the stellar (J-K) (and other indices) which, if uncorrected, results in an underestimate of the T$_{\rm eff}$. Photometry by Alksnis et al. (2002) describes how infrared magnitudes respond as the star varies. For a change at R of about 3 magnitudes, changes in magnitudes in L and M are only about 0.2 and -0.3, respectively, but considerably greater at J (2.1), H (1.7) and K (1.0). Although these changes indicate that photometry at M is dominated by the shell's quasi-constant contribution, precise attribution at H and K of the shell's contribution remains uncertain, even had we access to contemporaneous multi-color photometry across the infra-red. A shell continuum of about 35\% contribution to the K band will dilute DY Per's CO lines relative to the strongest lines in TX Psc and other carbon stars but since this contribution must rather similarly affect the spectrum across the entire K band, the CN lines could  be similarly  weakened relative to their strength in TX Psc.  Thus, the intervals shown in Fig. 1 suggest the DY Per's shell is not an important contributor to the star's K band.

\section{Chemical clues to DY Per's identity }

Clues to DY Per's identity are sought from its IGRINS spectrum as well as from  published accounts of its spectrum at shorter wavelengths. Is this cool C-rich variable related to the H-deficient HdC and RCB stars  or is it an unusually active mass-losing variant of a normal carbon AGB giant? Or, perhaps, DY Per is of  an entirely novel origin. Our analysis of the IGRINS spectrum is focused on two potential contrasting distinguishing marks between the H-deficient stars and normal carbon giants: their  atmospheric $^{16}$O/$^{18}$O ratio and the hydrogen-deficiency. The O isotopic ratio is obtainable directly by matching the key portion of the observed spectrum with  synthetic spectra computed from our molecular line lists and model atmospheres. The result is clear: DY Per is marked by the low $^{16}$O/$^{18}$O ratio, the striking signature of the  H-deficient RCB and HdC stars, and not the high ratio  found among normal carbon AGB giants.  In contrast,  DY Per's H-deficiency  can not be closely set  from either our K-band spectrum or published discussions of optical spectra. Presence of hydrogen in DY Per's atmosphere is revealed by HF vibration-rotation 1-0 lines in the K-band but the strength of HF lines may depend on the combination of the H and the F abundances. The expected  H-deficiency may possibly  be set by the condition that DY Per's F abundance should not exceed the  maximum  F abundance of H-deficient RCB and extreme He stars where their F abundance is set from optical spectra possessing atomic fluorine lines. 

Our principal tool for extracting abundance data from the  IGRINS spectra is to fit  DY Per's observed spectrum  with a synthetic spectrum convolved with the instrumental profile. Synthetic spectra are computed from a model atmosphere drawn from a set centered on  T$_{\rm eff} = 3000$ K , a gravity $\log g = +0.5$,  a metallicity  [M] of solar,  a microturbulence of 7 km s$^{-1}$. Three different  input H-abundances of 12.0 (normal = solar), 10.5 and 9.5 were available for each point on the grid.  The spectrum provided by TURBOSPECTRUM was convolved with a macroturbulence of 6 km s$^{-1}$ and the IGRINS instrumental profile before tested against the  observed spectrum.

\subsection{Hydrogen content}

Alksnis's (1994) identification from photometry of DY Per as a RCB variable  clearly calls for spectroscopic confirmation  of the star's H-deficiency. A first suggestion of H-deficiency came from Keenan \& Barnbaum (1997) who remarked that a `moderate hydrogen deficiency' may be present for DY Per from comparison of 3.4\AA\ resolution spectrum of the CH G-band for DY Per, the HdC star HD 182040, and the R-type carbon star RV Sct. Their remark has been widely cited to confirm DY Per as a RCB variable but  the claim lacks rigor. 

The sole attempt at a quantitative estimate of DY Per's H-deficiency from the CH transition appears to be by Yakovina, Pugach, Pavlenko  (2009) who constructed a model atmosphere grid and assembled atomic and molecular line lists to construct synthetic spectra.  Their fit to an observed optical spectrum at  3.4\AA\ resolution did not involve the heart of the CH G-band around 4320 \AA\ where the observed spectrum had too low a S/N ratio but the much weaker $\Delta v = -1$ CH bands  about 4850 \AA\ where fortunately the overlapping bands of the C$_2$ Swan system are weaker but the CH contribution appears to be weaker than the set of CN Violet system lines. From comparisons of observed and their synthetic spectra, Yakovina et al. (2009) put the H/He number ratio in the broad range 1/9 $\leq$ H/He $\leq 9/1$, i.e., a H-deficiency ranging down a factor of 100 but with a solar H/He ($\sim 9/1$) ratio also providing an acceptable fit to the spectrum. A feature  of their fits to the observed spectrum  is that their derived H-deficiency seems correlated with the derived Fe-deficiency; if DY Per's  metallicity is near-solar  (see below)  DY Per has a quasi-normal H content.  Our synthetic spectra  around 4840 \AA\ show that the C$_2$ Swan $\Delta v = 0$ may dominate the observed spectrum  with the CN Violet lines on average stronger than the CH lines  for a normal H abundance; establishing a H-deficiency will be difficult. This suggestion is confirmed by Crawford et al.'s  (2023) observation that the CH G-band was undetectable in spectra of HdC stars ``even those with measurable Balmer lines".

For some carbon stars, it is possible to infer the presence of  H$\alpha$ absorption  contributing to their  spectrum dominated by  a rich array of CN and C$_2$ absorption lines. In other carbon stars, the spectrum around H$\alpha$ appears entirely due to the CN and C$_2$ lines. A valuable survey of  Balmer lines in carbon stars was provided by Barnbaum (1994). An instructive discussion of a possible H$\alpha$ contribution to carbon stars is provided by Smirnova (2012). When Balmer lines are present, either in emission or absorption,  the supposition is that they originate from either a shock traversing the photosphere or a stellar chromosphere.  Presence of H$\alpha$ most probably eliminates the possibility of serious H-deficiency.
%Conversion of a Balmer line's intensity to a H abundance may be deemed impossible. 
%Presence of H$\alpha$ likely suggests the host  carbon star  lacks the severe H-deficiecy of a RCB variable
Absence of H$\alpha$ is consistent with presence of a carbon giant but cannot be taken as a conclusive identifier of a RCB variable.

Za\v{c}s et al. (2005, Fig. 5) compared the H$\alpha$ profile in high-resolution spectra of DY Per and the N-type carbon star U Hya. Across the 30 \AA\ illustrated portion, their U Hya spectrum is very similar to Barnbaum's (1994) U Hya spectrum which in turn is almost identical to her spectrum of TX Psc. DY Per's spectrum is {\it very} different in that it is dominated by few flux peaks between which the many discrete but blended absorption lines in U Hya have merged (i.e., strengthened) to create a floor with a minimum relative flux of 0.1. This contrast between U Hya (and TX Psc) and DY Per results from strengthening in the latter of CN and C$_2$ lines - see Barnbaum's templates of these lines. Neither the spectrum of U Hya nor of DY Per at H$\alpha$ indicates a contribution in either absorption or emission from H$\alpha$ but, as noted above, this result cannot not serve as an indicator of H-deficiency. This statement is compatible with Za\v{c}s et al 's  (2005) remark that the region around H$\alpha$ (and H$\beta$) is too crowded ``to clarify the level of hydrogen deficiency'' in DY Per. But in contrast, Za\v{c}s et al. (2007, Fig. 10) from their comparison of just 5 \AA\ around H$\alpha$ of DY Per and carbon-rich giants propose the conclusion that DY Per enjoys a ``significant hydrogen deficiency''. We suggest  our and the group's 2005 conclusion is to be preferred to this later one.

Infrared emission from DY Per confirms the presence of hydrogen but an estimate of
the H-deficiency is not yet available. The 11.3$\mu$m and 12.7$\mu$m emission
attributable to polycyclic aromatic hydrocarbon (PAH) molecules
(Garc\'{i}a--Hern\'{a}ndez, Rao \& Lambert 2013) are present in DY Per but absent
from very H-deficient RCB stars. Intriguingly, the PAH features are also present in
spectra of three moderately H-deficient RCB stars: the two hot RCB stars DY Cen and
HV 2671 and the warm RCB star V854 Cen. Atmospheres of these RCB stars have
H-deficiencies of around 100-fold. Such a moderate H-deficiency may apply to DY Per 
(see below) but the chemistry of PAH production and emission in cool shells is not
understood yet and estimating H-deficiencies is a most uncertain procedure.

%Spectroscopic determination of the H abundance for a carbon giant is an uncertain process, see remarks above about Balmer lines. 
The H$_2$ quadrupole transition 1-0 S(0) at 2.2226 $\mu$m offers another possibility, if DY Per is not too seriously H-deficient. This and the S(1) line appear in the infrared spectra of many carbon stars at close to the predicted equivalent width at the predicted wavelength (Lambert et al. 1986). But Ohnaka, Tsuji \& Aoki (2000) note that the H$_2$ line in some carbon stars is velocity shifted and may be formed in ``a warm molecular envelope'' above the stellar photosphere. Lambert et al. (1986), Aoki, Ohnaka \& Tsuji  (1998) and Ohnaka, Tsuji \& Aoki (2000) each conclude that the atmosphere of a typical carbon AGB giant is not H-deficient. Our IGRINS spectrum of DY Per places the S(0) line in the red wing of a weak line.  (As expected, the S(0) line is absent from  the HdC HD 137613's IGRINS spectrum.) The strength of S(0) in a carbon star of DY Per's T$_{\rm eff}$ is such that a slight deficiency of hydrogen reduces the line below its detection limit. 

A future and more favorable opportunity to probe the H-deficiency of DY Per may be provided by high-resolution spectra of the CH (and NH) fundamental vibration-rotation lines  near 4 $\mu$m but correction for infrared emission from DY Per's circumstellar shell may be required. CH (and NH) lines near 4$\mu$m were exploited previously by Lambert et al. (1986) and Ohnaka, Tsuji \& Aoki (2000) in spectroscopic studies of a few carbon giants.  

Discussion of DY Per's H abundance is continued with  discussion of the HF vibration-rotation lines  (below).

\subsection{The $^{12 }$C/$^{13}$C ratio}

%Optical surveys for new RCB and DY Per-like variables  when supplemented by  low-resolution optical spectra mention that   features in the CN Red system or the C$_2$ Swan bands  identified with the $^{13}$C isotope are generally  evident in the spectra of DY Per variables but rarely seen in RCB variables. This distinction first noted by Alcock et al. (2001) is not necessarily a reflection of  differences in the $^{13}$C abundance between RCB and DY Per variables, although  it may have been claimed so.  The low-resolution spectra show the molecular bands to be stronger for the DY Per variables  thus increasing the detectability of a  $^{13}$C--feature, even without a change in the isotopic ratio. 

%Inspection of spectra  near 8000 \AA, a region dominated by lines of the CN Red system,  for DY Per  and three cool carbon stars with  a known $^{12}$C/$^{13}$C  ratio (Lambert et al. 1986)  led    Za\v{c}s  et al.  (2007)  to suggest that   DY Per has `a relatively high  $^{12}$C/$^{13}$C   ratio', say, $\sim 50$, a mid-range  value for either  carbon giants or probably  too for RCB variables. Thus,  DY Per's  $^{12}$C/$^{13}$C estimate  is unlikely  to  referee the issue of `carbon giants or RCB origin?' 

Investigations of the $^{12}$C/$^{13}$C ratio among carbon AGB stars and the H-deficient RCB and HdC stars indicate that the ratio spans overlapping ranges in the two groups of stars. This result effectively eliminates the likelihood that DY Per's $^{12}$C/$^{13}$C ratio may serve to suggest its identity. Our sample of $^{12}$C/$^{13}$C ratios for carbon stars are taken from analyses by Lambert et al. (1986) who determined estimates from the CN Red system's $\Delta \it{v}  = -2 $ bands and the CO first- and second-overtone bands. A majority of carbon star sample have a $^{12}$C/$^{13}$C ratio between 30 and 80 with a few $^{13}$C-rich -- J-type -- having the recognizable ratio $^{12}$C/$^{13}$C $\simeq 4$ for the CN-cycle run at equilibrium.

For warm RCB stars, the $^{12}$C/$^{13}$C ratio has been obtained from C$_2$ Swan bands, particularly with the 1-0 $^{12}$C$^{13}$C band head at 4744 \AA\ referenced to the 1-0 $^{12}$C$_2$ band with its head at 4737 \AA. Synthesis of both bands is complicated by the accidental presence of Fe\,{\sc i} lines at the heads. Hema, Pandey \& Lambert (2012) estimate the $^{12}$C/$^{13}$C ratio for warm RCBs with Swan bands. Synthesis of weak Swan bands, even from high S/N high-resolution spectra often result in a lower limit to the  $^{12}$C/$^{13}$C ratio. An interesting result is that the two minority RCBs (i.e., RCBs with abnormally high [Si/Fe] and [S/Fe] ratios) returned $^{12}$C/$^{13}$C ratio estimates from 3 to 6 for VZ Sgr and 8 to 10 for V CrA (see also Rao \& Lambert 2008). Majority RCBs with normal [Si/Fe] and [S/Fe] returned $^{12}$C/$^{13}$C ratios of about 20 or lower limits of 8 to 120. Data on the  $^{12}$C/$^{13}$C ratio among RCB stars could be expanded from high-resolution infrared spectra, particularly from the CO lines of those cool RCBs with K-band continuum undiluted emission from a dust shell. Unfortunately, the small K-band pieces of the previous PHOENIX spectra of cool RCBs did not covered useful spectral portions for the derivation of the $^{12}$C/$^{13}$C ratio (Garc\'{i}a-Hern\'{a}ndez et al. 2009, 2010).

A large $^{12}$C/$^{13}$C ratio for HdC stars was suggested long ago by Fujita \& Tsuji (1977) from high-resolution spectra providing CN Red system lines around 8000 \AA\ and  consideration of $^{12}$CN and $^{13}$CN lines of the same intensity. They found that the ratio was $> 100$ for HD 182040 and $> 500$ for HD 137613. Kipper (2002) obtained $^{12}$C/$^{13}$C $> 40$ for HD 137613 from synthesis of CN lines in the spectrum around 8000 \AA. Hema, Pandey,  Lambert  (2012) from analysis of the Swan bands report the ratio to be $> 100$ for HD 137613 and HD 175893 and slightly lower ($>60$) for HD 173409 and higher ($> 400$) for HD 182040. From K-band PHOENIX spectra, Garc\'{i}a-Hern\'{a}ndez et al. (2009)  put lower limits of $^{12}$C/$^{13}$C$>$10 for these three HdC stars. 

 DY Per's spectrum around the $^{13}$C$^{16}$O 2-0 R-branch band head was synthesized  for  $^{12}$C/$^{13}$C ratios of  4,  15 and 30.  This   selection of  $^{13}$C$^{16}$O lines is crossed by strong lines from 2-0 and 3-1 $^{12}$C$^{16}$O bands, CN and C$_2$ lines and very weak $^{12}$C$^{17}$O 2-0 lines.  The $^{13}$C$^{16}$O band head and individual lines certainly exclude the ratio $^{12}$C/$^{13}$C  =  4 and, hence, serious contamination by CN-cycled material.  A limit $^{12}$C/$^{13}$C $= 15\pm5$ is accepted from this preliminary search.  It seems unlikely that investigation of other portions of the  $^{13}$C$^{16}$O spectrum will improve the estimate of the ratio because across these portions lines from all CO isotopologues overlap. Detailed scrutiny of the spectrum short wards  of the 2-0  $^{12}$C$^{16}$O band (see Fig. 1)  might yield  a higher $^{12}$C/$^{13}$C ratio from either CN or C$_2$ lines, notably from the latter where the number densities n($^{13}$C$^{12}$C) = 2n($^{12}$C$_2$)/$r$ where $r$ = $^{12}$C/$^{13}$C.
 
 Although the distribution functions of $^{12}$C/$^{13}$C ratios are not  yet well defined for available  samples of   RCB and HdC stars, the functions  possibly span different ranges with HdC stars exhibiting  larger $^{12}$C/$^{13}$C ratios  than either carbon giants or the RCB stars.  A  more precise estimate for DY Per seems  unlikely to be the key to its identity.

\subsection{The isotopic oxygen ratios: $ ^{16}$O/$^{18}$O and $^{16}$O/$^{17}$O }

%fig2
\begin{figure*}
%\figurenum{3}
\includegraphics[angle=-90,scale=.6]{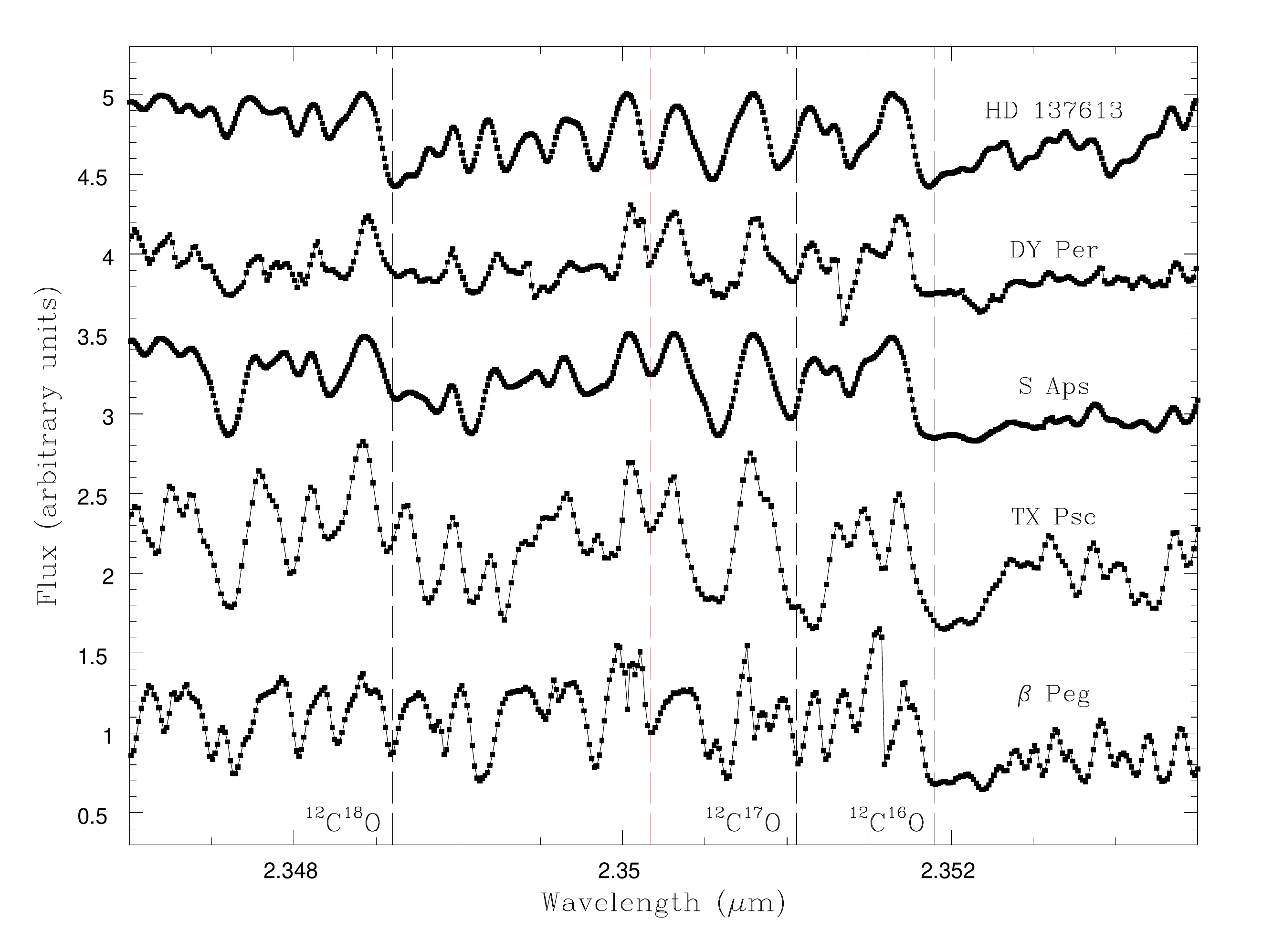}
\caption{The wavelength interval 2.3470$\mu$m to 2.3535$\mu$m including the R branch heads (denoted by vertical dashed lines and each degraded to longer wavelengths) for the 4-2 $^{12}$C$^{18}$O, 3-1 $^{12}$C$^{17}$O and 2-0 $^{12}$C$^{16}$O bands. The red dashed line identifies the R(43) 4-2 $^{12}$C$^{18}$O line, which appears unblended in each illustrated spectrum. The five stars whose spectra are shown on the same relative flux scale but displaced vertically are the HdC HD 137613, DY Per, the RCB S Aps, the carbon star TX Psc and the M giant $\beta$ Peg. \label{fig2}}
\end{figure*}

%fig3
\begin{figure*}
%\figurenum{4}
\includegraphics[angle=-90,scale=.6]{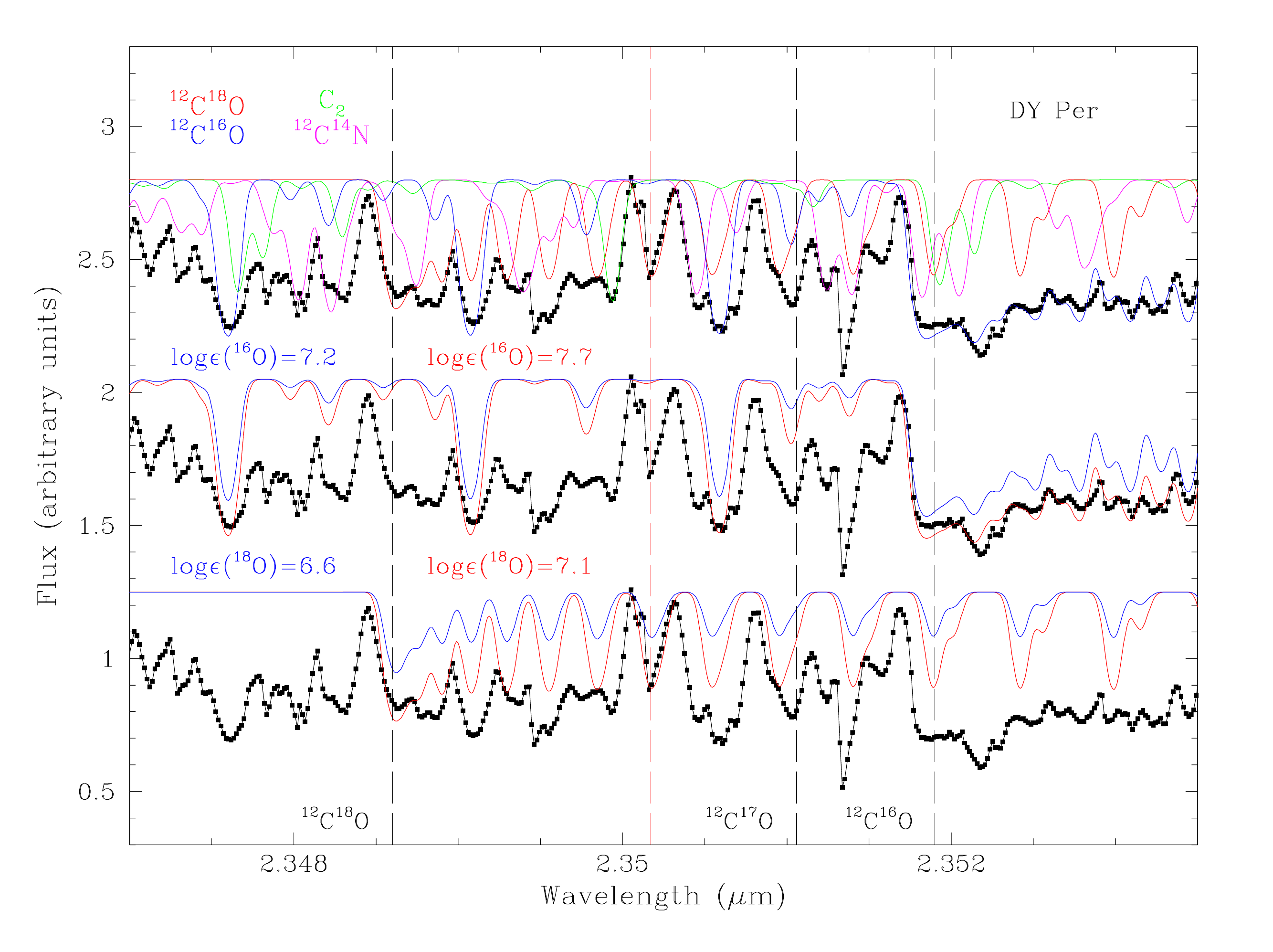}
\caption{The wavelength interval 2.3470$\mu$m to 2.3535$\mu$m including the R branch heads (denoted by vertical dashed lines and each degraded to longer wavelengths) for the 4-2 $^{12}$C$^{18}$O, 3-1 $^{12}$C$^{17}$O and 2-0 $^{12}$C$^{16}$O bands. The red dashed line identifies the R(43) 4-2 $^{12}$C$^{18}$O line, which appears unblended. The observed DY Per spectrum is illustrated three-fold. At the top, fitted synthetic spectra provided by the molecules $^{12}$C$^{16}$O (blue), $^{12}$C$^{18}$O (red), $^{12}$C$^{14}$N (magenta) and C$_2$ (green) are fitted to the DY Per spectrum. (Contributions from $^{12}$C$^{17}$O are not shown because these lines are unimportant (see text).)  In the middle, synthetic spectra for oxygen $^{16}$O abundances $\log \epsilon$($^{16}$O) = 7.2 and 7.7 are shown. At the bottom, synthetic spectra for oxygen $^{18}$O abundances $\log \epsilon$($^{18}$O) = 6.6 and 7.1 are shown. We note that the displayed synthetic spectra are constructed with a model atmosphere with T$_{\rm eff} = 3000$ K, $\log g = +0.5$, [M]=0.0, microturbulence = 7 km s$^{-1}$, and an input H-abundance of 10.5 but the derivation of the $^{16}$O/$^{18}$O is independent of the model atmosphere used (see text). \label{fig3}} 
\end{figure*}

The $^{16}$O/$^{18}$O ratio appears to be a striking discriminant between  N-type AGB carbon giants and typical RCB-variables and HdC stars. Determination of oxygen isotopic ratios for RCB and HdC stars is dependent on these H-deficient stars having a sufficiently cool temperature to provide CO lines in their infrared spectrum dominated by that of the  star with no or minor contamination from a dusty cold circumstellar  shell. CO bands do not occur in spectra of the warmest RCB stars and certainly not the extreme helium stars and, thus, their O isotopic ratios remain unknown. %Presently, available determinations of the $^{16}$O/$^{18}$O ratio demonstrate the clear distinction between ranges of isotopic ratios among N-type carbon stars and among RCB-variables and HdC stars sought as indicating DY Per's closest relatives. 

Isotopic oxygen ratios in carbon stars are provided in independent analyses by Harris et al. (1987) and by Abia et al. (2017) of  CO lines in the high-resolution spectra acquired by Lambert et al. (1986). These analyses of oxygen isotopic ratios are in fair agreement. Abia et al. (2017) give the mean ratios as $^{16}$O/$^{17}$O $=1057\pm460$ and $^{16}$O/$^{18}$O $=870\pm280$, which are lower than Harris et al.(1987)'s values by $-225\pm500$ and $-500\pm620$, respectively. Equivalently, the ratio of the Abia/Harris r estimates is on average 0.92 and 0.72 for $^{16}$O/$^{17}$O and $^{16}$O/$^{18}$O, respectively. Abia et al. (2017) find that the $^{16}$O/$^{18}$O across the sample runs from $500\pm200$ to $1600\pm150$. The $^{16}$O/$^{17}$O shows a similar range. These isotopic ratios are about what are predicted for carbon AGB giants created by the third dredge-up in low-mass stars. 

Discovery of the remarkable $^{16}$O/$^{18}$O ratios for HdC and RCB stars from direct
inspection of low-resolution spectra took advantage of the displacement of the
$^{12}$C$^{18}$O 2-0 R-branch band head at 2.3486 $\mu$m from the  $^{12}$C$^{16}$O  4-2
R-branch at 2.3519 $\mu$m (Clayton et al. 2005, 2007; Garc\'{i}a-Hern\'{a}ndez et
al. 2010). Between these two band heads is the $^{12}$C$^{17}$O 3-1 R-branch band head
which was far from prominent signifying that $^{16}$O/$^{17}$O $>>$  $^{16}$O/$^{18}$O.
The region 2.348 to 2.354 $\mu$m includes not only lines from these CO bands but also
lines from the 2-0 and 3-1 $^{12}$C$^{16}$O bands, and the 2-0 $^{13}$C$^{16}$O band.
Crucially, additional lines across this region come from the CN Red system, the C$_2$
Phillips and Ballik-Ramsay systems. When the $^{18}$O abundance is high,
$^{12}$C$^{18}$O lines from bands other than the 2-0 are detectable but these lines
being at longer wavelengths fall amongst an increasing dense multi-band mix of
$^{12}$C$^{16}$O and  $^{13}$C$^{16}$O lines, frustrating accurate determination of the
$^{16}$O/$^{18}$O ratio. 

An impression of DY Per's $^{16}$O/$^{18}$O ratio is provided by the spectra in Fig. 2 of four carbon-rich stars in order of increasing $^{16}$O/$^{18}$O from top to bottom with the M giant $\beta$ Peg's spectrum at the bottom. Vertical lines mark the locations of the R-branch heads of the 2-0 $^{12}$C$^{18}$O, 3-1 $^{12}$C$^{17}$O, and the 4-2 $^{12}$C$^{16}$O, all degraded to longer wavelengths. HdC HD 137613 clearly has a $^{16}$O/$^{18}$O $\sim 1$ as shown by the nearly equal depths of the 2-0 $^{12}$C$^{18}$O and the 4-2 $^{12}$C$^{16}$O band heads and also by the clear rotational structure of the 2-0 $^{12}$C$^{18}$O band. RCB S Aps not only has a 2-0 $^{12}$C$^{18}$O head weaker than the 4-2 $^{12}$C$^{16}$O head indicating that its $^{16}$O/$^{18}$O ratio ($16\pm4$) is larger than for HD 137613 but the profile of S Aps's  2-0 $^{12}$C$^{18}$O band is interrupted by strong 3-1 R branch $^{12}$C$^{16}$O (and other lines) lines. This H-deficient pair are of higher $T_{\rm eff}$ (5400 K) than either DY Per or TX Psc ($T_{\rm eff} = 3030$ K). DY Per appears to offer a spectrum with the 2-0 R branch $^{12}$C$^{18}$O closer in strength to its 4-2 R branch $^{12}$C$^{16}$O than their relative strength in S Aps. In S Aps, the region shortward of the 2-0 $^{12}$C$^{18}$O head is seen to be richer in contamination from CN and C$_2$ lines which are present across the K band and call for careful synthesis before assessing a carbon star's $^{16}$O/$^{18}$O ratio. Mere inspection of the spectra of TX Psc (T$_{\rm eff} \sim 3030$ K) and $\beta$ Peg (T$_{\rm eff} \sim 3270$ K) does not so readily betray their $^{16}$O/$^{18}$O ratios which are the typical high values of giants ($^{16}$O/$^{18}$O $\sim 600\pm370$ for TX Psc (Abia et al. 2017) and $\sim 830$ for $\beta$ Peg (Lebzelter et al. 2019). This seeming betrayal arises in part from contamination by other lines, principally from the CN Red system in the case of TX Psc, contributions from $^{13}$C$^{16}$O lines in the case of $\beta$ Peg, and the necessity to adjust impressions for saturation of strong $^{12}$C$^{16}$O lines, and their high $^{16}$O/$^{18}$O ratio.

Before discussing quantitative estimation of the $^{16}$O/$^{18}$O ratio for DY
Per, we note that Fig. 2's  red dashed line denoting the  2-0
$^{12}$C$^{18}$O R43 line appears with a similar strength in DY Per, S
Aps and TX Psc, and also in our published PHOENIX spectra of RCBs U Aqr and V1783 Sgr
but is unsurprisingly much stronger in HD 137613 with its high $^{18}$O
abundance. The `red' line appears stronger in $\beta$ Peg because this M
giant has an appreciable $^{13}$C content ($^{12}$C/$^{13}$C = 9; Lebzelter, Hinkle, Straniero et
al. 2019), also  the 2-0 $^{12}$C$^{18}$O line is blended with the 2-0 R69
$^{13}$C$^{16}$O line. The R68 $^{13}$C$^{16}$O line (and similar lines) is also
clearly present in $\beta$ Peg. (The illustrated $\beta$ Peg spectrum in the
neighborhood of the `red line' and a few places elsewhere shows abrupt local 
changes of intensity which arise largely  from incorrect cancellation for the
telluric lines.) 

Subject to correction for the $^{13}$C$^{16}$O blend (and for as yet unidentified blends), the `red' line provides an estimate of the atmosphere's $^{12}$C$^{18}$O abundance. This estimate should be matched with lines of comparable or weaker intensity from $^{12}$C$^{16}$O in order to obtain the $^{16}$O/$^{18}$O ratio with minimal sensitivity to curve of growth effects but an inevitable sensitivity to effective temperature. In HD 137613 and similar stars, there are opportunities to isolate unblended or only weakly-blended lines of $^{12}$C$^{18}$O and $^{12}$C$^{16}$O but for carbon stars like TX Psc where $^{18}$O (and also $^{17}$O) is a trace isotope and blending CN and C$_2$ lines are a serious contaminant unblended $^{12}$C$^{18}$O lines are rare and weak $^{12}$C$^{16}$O lines uncommon. Differences in excitation potential between the CO varieties also enter into a quantitative comparison. 

Extraction of the $^{16}$O/$^{18}$O ratio for DY Per is illustrated by Fig. 3  where the IGRINS spectrum from Fig. 2 is repeated. Superimposed on the observed spectrum and distinguished by different colors are predicted contributions from not only $^{12}$C$^{16}$O and $^{12}$C$^{18}$O but also from $^{12}$C$^{14}$N and C$_2$. The individual molecular contributions are displayed in order to show the complexity of the observed spectra and they correspond to the initial guesses for the CNO abundances. Several iterations using specific spectral regions dominated by $^{12}$C$^{14}$N ($\sim$2.251 $\mu$m) and $^{12}$C$^{16}$O ($\sim$2.343 $\mu$m), among others, were needed to obtain the final elemental abundances of C/N/O=9.4/8.6/7.9 with the T$_{\rm eff} = 3000$ K, $\log g = +0.5$, and [M]=0.0 model with an input H-abundance of 10.5 and microturbulence = 7 km s$^{-1}$).

The synthetic spectrum fitted to the IGRINS spectrum corresponds to $^{16}$O/$^{18}$O $=4\pm1$. As noted above, the appearance of the spectrum indicates that the $^{12}$C$^{16}$O and $^{12}$C$^{18}$O contributions may be estimated almost independently and this is explored in the lower panels in Fig. 3. The middle panel repeats the synthesis but with two values for the $^{16}$O abundance and the bottom panel shows the synthetic spectra for the $^{12}$C$^{18}$O contribution with the $^{18}$O abundance assigned two values. We stress that our measurement of the $^{16}$O/$^{18}$O is robust and  insensitive to the model atmosphere adopted. For example, very similar results are obtained for [M]=0.0 3000 K models with a H=9.5 and 12 ($=4\pm1$), a metal-poor [M]=-1.0 3500 K model with H=9.5 and 10.5 ($=6\pm2$), and even with a normal sequence model atmosphere constructed for the stellar parameters of the C-rich AGB star TX Psc ($=5\pm3)$.  This low $^{16}$O/$^{18}$O ratio was anticipated by Bhowmick et al.'s (2018) analysis of an infrared spectrum at the low resolution $R$ $\sim$ 900 which gave $^{16}$O/$^{18}$O $\geq 4\pm0.2$. 

Discovery of high $^{18}$O abundances from inspection of low-resolution spectra of HD 137613 and friends showing an unusual depth of the R branch of the 2-0 $^{12}$C$^{18}$O band relative to R branch of the 4-2 $^{12}$C$^{16}$O band but was not accompanied by remarks on the $^{17}$O abundances (Clayton et al. 2005, 2007). The 3-1 $^{12}$C$^{17}$O R branch head at 2.3511 $\mu$m falls between these two heads and just short of the 4-2 $^{12}$C$^{16}$O head in an interval dominated by several strong lines such that the strength of $^{12}$C$^{17}$O lines was presumed better examined from high-resolution spectra. Detection of $^{12}$C$^{17}$O lines is likely best attempted from its 2-0 R branch, as first noted by Maillard (1974) for M giants, and  exploited  by Harris et al. (1987) and Abia et al. (2017) for  carbon giants. Inspection of the spectra shows that the $^{12}$C$^{17}$O lines 2-0 R25 to R33 appear absent in both HD 137613 and DY Per, notably  lines with J= 27, 32 and 33. For DY Per, the synthetic spectra provide the lower limit 
$^{16}$O/$^{17}$O$\geq$80), a lower limit similar to those derived for HD 137613
and other RCB and HdC stars.

DY Per's high $^{18}$O abundance places this  C-rich giant among the H-deficient RCB and HdC stars and quite apart from the carbon AGB giants. Karambelkar et al. (2022) analyzed low-resolution (mostly at $R$ $\sim 3000$) spectra of the CO bands. DY Per and related variables were not observed. Our $^{16}$O/$^{18}$O $ =4$ for DY Per is within the range reported by the authors for their extensive sample of H-deficient giants.  Karambelkar et al. (2022) claim that the HdC  and RCB stars have ``different oxygen isotope ratios":  six of the HdC stars have $^{16}$O/$^{18}$O $< 0.5$ but in contrast 28 of the 33 RCB stars have $^{16}$O/$^{18}$O $> 1$ but 12 RCB stars are determined to have $^{16}$O/$^{18}$O $>  50$  and even 8 of this sample have $^{16}$O/$^{18}$O $> 500$.  It is surely, as Karambelkar et al. (2022) note, vital to confirm the distribution function for $^{16}$O/$^{18}$O  among HdC and RCB stars from high-resolution infrared spectra  but these initial differences may point to  a difference  in  origins of HdC and RCB stars. Intriguingly, extension of this hint to DY Per variables could yield insight into a possible relation between these variables and the HdC/RCB stars.

%fig4
\begin{figure*}
%\figurenum{5}
\includegraphics[angle=-90,scale=.6]{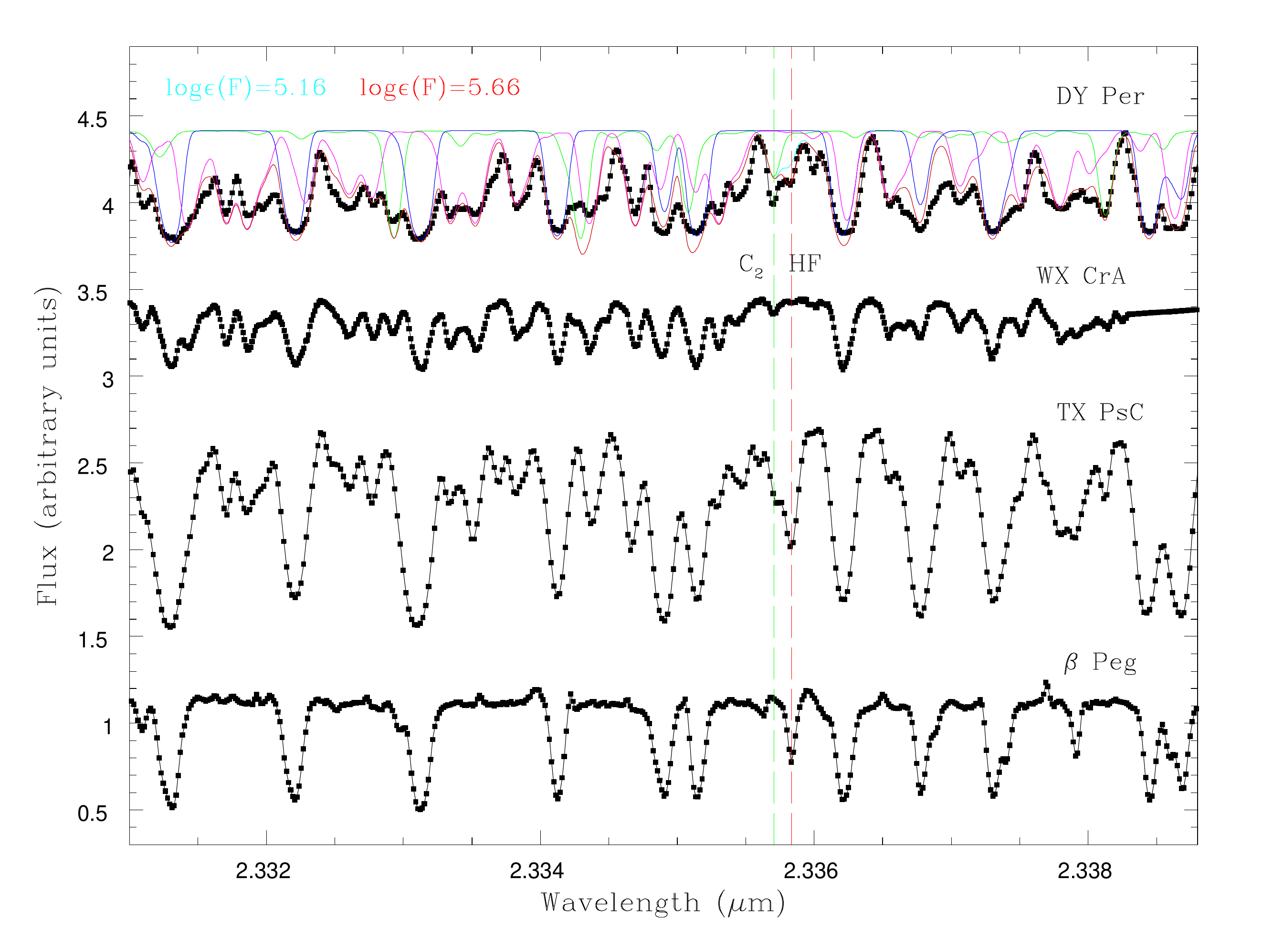}
\caption{The wavelength interval 2.3310$\mu$m to 2.3390$\mu$m, including the HF 1-0 R(9) line at 2.33583$\mu$m identified by the red dashed vertical line, is shown here (from top to bottom) for DY Per, the HdC-RCB WX CrA, the carbon giant TX Psc and the M giant $\beta$ Peg. In spectra of carbon--rich giants, the HF line is accompanied by a C$_2$ line  at a slightly shorter wavelength and identified by the green dashed vertical line. On DY Per's spectrum are superimposed contributions to the synthetic spectrum from $^{12}$C$^{16}$O (blue), $^{12}$C$^{14}$N (magenta), and C$_2$ (green). The global synthetic spectra for the model atmosphere with T$_{\rm eff} = 3000$ K, $\log g = +0.5$, [M]=0.0, and H=10.5 (microturbulence = 7 km s$^{-1}$) and different F abundances of log($\epsilon$)=5.2 (cyan) and 5.7 (red) are also shown. \label{fig4}}
\end{figure*}

% Please remove the label HF at top left and the predicted red Gaussian in the DY Per synthetic spectrum.
% We should check contribution to HF 1-0 R(9) from 12C17O 2-0 R(25) which is 0.208 cm-1 to higher wave numbers
% VERTICAL SCALE - FLUX (arbitrary units) - IS THE SAME IN UNITS OF LOCAL CONTINUUM FOT EACH STAR??

\subsection{The Fluorine abundance}

Fluorine's abundance is a potential indicator of DY Per's heritage. Carbon AGB giants betray F's presence through vibration-rotation 1-0 lines of HF in their K-band spectrum.  %RCB-variables and HdC stars do not show HF lines presumably on account of their warm temperatures and possibly the HF density is so reduced by the severe H-deficiency. 
Spectra of the warmest RCB-variables and extreme He stars show variously lines of  F\,{\sc i} and F\,{\sc ii} thus providing a measure of the F abundance among these very H-deficient stars. Through a, careful interpretation of the HF lines in DY Per's spectrum,  it may be possible to highlight DY Per's heritage. 

Fluorine in a cool C-rich giant is detectable through vibration-rotation 1-0 lines of HF in the K-band spectrum. J\"{o}nsson et al. (2014) review 1-0 vibration-rotation spectrum. The 1-0 R(9) line at 2.33583$\mu$m appears unblended in spectra of late-type stars. Fig. 4 shows an interval around this HF line in four giants: DY Per, the HdC WX CrA, TX Psc, and the M giant $\beta$ Peg. The HF line is marked by the red dashed line. The green dashed line marks the location of a nearby C$_2$ line. The R(25) $^{12}$C$^{17}$O 2-0 line is blended with the C$_2$ line but is not a serious contributor.  This C$_2$ line appears blended with an  unidentified line. 

Inspection of Fig. 4 confirms expectations about the spectra of the four giants:  $\beta$ Peg offers the cleanest spectrum with the HF line, WX CrA has a weak C$_2$ blend but no detectable HF line, the CN and C$_2$ lines between the prominent $^{12}$C$^{16}$O lines in TX Psc and DY Per (also contributing in WX CrA) are a more severe contaminant for DY Per than TX Psc because, as shown by Fig. 1, all CO lines are weaker in DY Per than in TX Psc.  This wavelength interval  appears not to be populated by lines from $^{12}$C$^{18}$O and $^{13}$C$^{16}$O. 
%Curiously, the HF R(9) line may be  free of blending by a  CO, CN or C$_2$ line. 

Fig. 4 suggests  that a severe H-deficiency and the higher T$_{\rm eff}$ ensure that the HF line is absent from WX CrA. Our PHOENIX observations (Garc\'{i}a-Hern\'{a}ndez et al. 2010, Fig. 2) show the HF line also absent from the HdC HD 137613 and also the observed cool (T$_{\rm eff} \sim 5500$ K) RCBs -- all with strong stellar CO bands  and presumably severe H deficiencies -- ES Aql, U Aqr, SV Sge, and S Aps. Our conclusion is that the absence of the HF line in these stars may be set primarily by their severe H-deficiency and not offset by an F overabundance is consistent  with a compilation of H and F abundances  for EHe stars and warm RCB variables (Jeffrey et al. 2011): mean values for abundances (H,F) are (6.8, 6.7) and (6.2,6.8) for EHe and RCB samples, respectively with the H abundances covering a range of  at least 3 dex and several RCB and EHe stars with [F/H]  at positive values. Exceptional  stars exist including the RCB V854 Cen with (H,F) = (9.9,$<5.7$)  and DY Cen at H=10.7 and lacking a F estimate. 

Published analyses of HF lines in TX Psc and $\beta$ Peg show that these giants have their expected F abundance -- assuming a normal H abundance --   
e.g., [F/Fe] $\sim 0.0$. Abia et al. (2010, 2015, 2019) report that a carbon star's  F abundance increases with its $s$-process enrichment approximately as expected for the third dredge-up phases. Since TX Psc is an $s$-process enriched carbon star, its F abundance of $\log \epsilon$ (F) $\simeq 4.7\pm0.1$ may be considered normal. The HF lines from Galactic M giants including $\beta$ Peg were analyzed by Guer\c{c}o et al. (2019) to obtain the star's $\log \epsilon$ (F) $= 4.04\pm0.05$ and [F/Fe] $= -0.10$. The fact that TX Psc's HF line is stronger than $\beta$ Peg's results primarily from three factors: the M giant has a $T_{\rm eff}$ nearly 600 K hotter than the carbon giant resulting in less of the hydrogen tied up in H$_2$ and it lacks the F enrichment from the third dredge up. 

Inspection of Fig. 4 shows that DY Per's HF line is weaker than in TX Psc. But this weakening  may reflect the H-deficiency  of DY Per. A H-deficiency affects both the line and the continuous opacities.
%Application of a model from the H-normal sequence (as was applied to TX Psc) returns for DY Per the F abundance $\log \epsilon$ (F) $= 2.8$ -- an unexpectedly low F abundance  for a H-deficient or a normal star. Adoption of H-deficiency reduces the HF density and, thus, the line opacity for a given F density. 
With H atoms  replaced by He atoms the  free-free infrared absorption of a He$^-$ ion is less effective than of a H$^-$ ion, the continuous absorption is  reduced but by a smaller factor than the line opacity.  Additionally, formation of H$_2$ molecules reduces the partial pressure of H atoms in DY Per and stars of similar temperatures.  Our  3000 K solar metallicity model atmosphere with  a normal H (=12) abundance fits the HF line with F  $= 5.2$ or [F/Fe] = +0.6 which is hardly an exceptional result given our lack of detailed information on the star's metallicity and its $s$-process enhancement. The equivalent model with H=10.5 (i.e., He/H = 10 not the solar He/H = 0.1), shows that the fit to DY Per's HF line is achieved with a F abundance of $\log \epsilon$ (F) $= 5.7$ ([F/Fe]=+1.1; see Fig. 4). The model with H=9.5 requires a F abundance of $\log \epsilon$ (F) $= 5.9$ ([F/Fe]=+1.3. These results are just within the spread of F overabundances found from atomic lines measured in EHes and warm RCB-variables but notably the H-deficiency is far smaller than the typical H-deficiency of 5 to 6 dex for EHe and RCB stars. Model atmospheres constructed for greater H-deficiencies may not yield much greater F abundances because the opacities are already dominated by He$^-$. 

  % If the typical H-deficiency applies to DY Per, the star would seem to have  a {\it very} high F abundance. Adoption of a metallicity far below solar with the expected sub-solar initial F abundance would require a greater H-deficiency but the typical H-deficiency seems out of reach.

% An F overabundance within the spread of F overabundances observed in EHes and warm RCB-variables would demand the introduction of even  more H-deficient models or alternatively model atmospheres with lower  T$_{eff}$ (e.g., 3500 K) and/or smaller metal abundances (e.g., [M/H]=-1.0). 

\subsection{C, N and O abundances}

Elemental C, N and O abundances are based on familiar conditions drawn from molecular equilibrium: the abundance drawn from the fit to C$_2$ lines  is a measure of (C-O)$^2$ , the CN lines sample the  quantity (C-O)N$^{0.5}$ and the CO lines in this C-rich atmosphere measure the  O abundance. The CO density is determined by the O abundance on account of the dominance of CO and N$_2$ in the molecular equilibrium.  Our abundance analysis  yields a striking difference in C, N, and O abundances between DY Per and carbon giants as represented by TX Psc. Given that DY Per and TX Psc have similar atmospheric parameters,  the evident weaker CO lines in DY Per relative to TX Psc but the not dissimilar strengths of C$_2$ and especially  the CN lines is a strong hint that DY Per is  O-poor relative to the carbon giant TX Psc. In turn, TX Psc is  representative of the carbon giants analyzed by Lambert et al. (1986). This inference is confirmed by fits of synthetic spectra to typical K-band windows sampling C$_2$, CN, and CO lines. 

Our adoption of the 3000 K and solar metallicity model with moderate H-deficiency represented by H = 10.5 provides the elemental abundances:
[C]=0.8, [N ]=0.6  and [O]=-0.9, which may be compared directly with results compiled by Jeffery, Karakas \& Saio  (2011) for samples of EHe and RCB stars.  DY Per's  [C]   and [N] are within the  range shown by these H-deficient stars for [Fe] = 0  but the [O]   is  close to the lowest values shown by EHe and RCBs.  Overlap between DY Per's C, N, and O abundances  and  samples of EHe and RCBs is improved if DY Per is somewhat metal-poor. 

Infrared photometry has likened  DY Per (and associated variables)  to carbon N-type giants.   Such a correspondence  is contradicted by our demonstration that DY Per's $^{16}$O/$^{18}$O ratio is   thoroughly unrepresentative of values reported for carbon giants. Now, DY Per's C/O ratio  of about 30  stands in sharp disagreement with values C/O $\sim 1$ found for carbon giants  (Lambert et al. 1986: Ohnaka, Tsuji \& Aoki 2000).  These correspondences point to a conclusion that DY Per's origin  lies not along an exceptional evolutionary path of a carbon giant but within the  origin of RCB  variables., i.e. by a white dwarf merger and subsequent mass loss and evolution to a hotter star. 

\section{DY Per's heritage?}

DY Per was isolated among the family of C-rich long period variables by Alksnis's (1994) proposal that its deep declines showed it `probably belongs to the RCB-type variables'.  Exploration of this proposal has been the focus of our investigation. Photometric surveys and their follow-up have shown that DY Per and newly discovered DY Per  variables have IR colors representative of carbon AGB giants but atypical of RCB variables. DY Per variables, as carbon giants at maximum light have lower effective temperatures than the coolest RCB variables, also at maximum light. Ability to determine the severe H-deficiency is, as we have highlighted, greatly impaired at a carbon giant's temperature and, thus, a defining marque of a RCB variable is unavailable. Our demonstration that DY Per has the low $^{16}$O/$^{18}$O ratio now (apparently) common among HdC and RCB variables presenting CO vibration-rotation bands in their spectra endorses the DY Per-RCB link. Our analysis of DY Per's K-band spectrum shows that the  star's C, N, and O abundances overlap well with the abundances of RCB variables and EHe stars but not with the reported compositions of carbon AGB giants.  These abundance differences between DY Per and carbon giants support the suggestion that DY Per  is likely  not  an exceptional  descendant of the common carbon giants.  This suggestion  echoes  Warner's (1967)  prescient discussion following his bold early abundance analysis of optical photographic spectra and  a suggestion reinforced by  discovery of abundant $^{18}$O in atmospheres of RCB-variables and HdC stars.   Association of  DY Per  rich in $^{18}$O with a carbon giant has to accept the challenge that  H-burning in the  giant's interior results in a $^{14}$N-rich zone which with the onset of He-burning is converted to  $^{18}$O via $\alpha$-capture. Appearance of abundant $^{18}$O in the atmosphere requires not only the avoidance of internal conversion of $^{18}$O to $^{22}$Ne by $\alpha$-capture but, in the absence of large-scale convective currents between the $^{18}$O-rich interior and the atmosphere, serious (unanticipated) mass-loss by the AGB giant is a necessity to expose the $^{18}$O-rich hot interior layer as part of the cool atmosphere of a H-deficient giant. This is not to deny  that  theoretical  imagination  is incapable of devising  schemes for  severe mass loss and deep mixing  capable of  producing a DY Per-like giant.  A cautionary  note is  that we have analyzed  just one DY Per variable, not established definitively its H-deficiency  nor that it is (or was) a single star.

If extraordinary evolution of a carbon giant is not easily seen as part of DY Per's heritage,  the likely alternative  involves a  merger of a He white dwarf with a C-O white dwarf leading to  theoretical RCB variables and EHe stars  in their observed locations in the Hertzsprung-Russell (H-R) diagram and with about their predicted compositions, notably the low  $^{16}$O/$^{18}$O ratios.  Birth of RCB variables in the merger of a He white dwarf onto a C-O white dwarf was explored broadly by Webbink (1984). Quantitative exploration of the properties including the composition of the merger product poses computational challenges and realistic representation of the physics at all phases from beginning of the merger of the He white dwarf onto the C-O white dwarf through to the emergence of the RCB product. The challenges in modeling may be appreciated by reading a sample of recent studies involving various He and C-O white dwarf pairs of solar and sub-solar metallicities:  Menon et al. (2013), Zhang et al. (2014), Crawford et al. (2020), Munson et al. (2021), and Munson, Chatzopoulos \& Denissenkov (2022). In broad terms, the RCB variables emerging from a merger may match the observed RCB and EHe stars in terms of H, C, N, O  and F abundances. A low $^{16}$O/$^{18}$O ratio is predicted by many models but can not be assumed to be a guaranteed prediction across a family of merger models.

RCB variables are low mass giants with effective temperatures from about 4000 K to 8000 K with luminosities $\log L/L_\odot  \sim 3.6$.  At temperatures below about 3500 K, the C-rich  giant is a spectroscopist's nightmare;  cool RCB stars  such as DY Per likely remain undiscovered spectroscopically. A few hotter RCB stars (e.g., DY Cen with a present T$_{\rm eff} \sim 25000$ K; Jeffery, Rao \& Lambert 2020) are known. The high temperature limit is extended and to lower luminosities by EHe stars with similar compositions to RCB stars. This evolution of the supergiants to increasing temperature is driven largely by mass-loss by the RCB as it evolves to higher temperatures. Evolutionary tracks in theoretical H-R diagrams displaying  white dwarf merger products predict giants with composition changes at these observed RCB luminosities and effective temperatures. DY Per's absolute luminosity is close to the range expected for merger products but its T$_{\rm eff}$ appears cooler than presently provided by  a representative sample of merger studies using different prescriptions to model the merger process: presently available  calculations, for example, Zhang et al. (2014) and Munson, Chatzopoulos \& Denissenkov (2022) predict  merger products  at least  about 1000 K warmer than the effective temperature of  DY Per.  But more detailed accounting of DY Per's structure including the opacity of the atmosphere and envelope may reduce the effective temperature.

\section{Concluding remarks}

Alksnis's (1994) bold suggestion from long-term photometry that the carbon star DY Per is a cool example of a RCB variable is supported but not unerringly proven by our demonstration that DY Per shares with the RCB variables and their undoubtedly close relatives the HdC stars the strikingly low $^{16}$O/$^{18}$O ratio which suggests these stars owe their origin to the merger of a He with a C-O white dwarf. H-deficiency, a signature of RCB, HdC and EHe stars, has yet to be proven for DY Per; the suspicion of H-deficiency for DY Per is encouraged but by no mans proven by the demonstration that a $^{16}$O/$^{18}$O ratio and a F overabundance are generally coupled and the HF line in DY Per is consistent with a moderate to a strong H-deficiency and the typical F overabundance for a RCB and EHe star. 

%Expanded examination of an association between RCB variables and the DY Per-variables discovered from Galactic and Magellanic Clouds surveys for RCB-like variables is clearly essential. Initial key insights are guaranteed from K-band spectroscopy and estimates of the $^{16}$O/$^{18}$O ratio and the strength of the HF lines. Additional information will be provided from determinations of $^{12}$C/$^{13}$C and $^{16}$O/$^{17}$O ratios. A key missing uncertainty in securing the identity of DY Per and DY Per variables as  close relatives of the RCB-variables is the proof of a serious H-deficiency. Here, the hope may be, as noted above, acquisition of the 4 $\mu$m spectrum and a search for the CH fundamental vibration-rotation lines. Detection of these CH lines would point to a lack of a serious H-deficiency but a quantitative demonstration of a H-deficiency similar a RCB variable by about 6 dex seems most unlikely. 

An emphasis on expanded  observational studies of DY Per and DY Per variables  should not obscure the need for continued theoretical work on the white dwarf merger process. The above challenge is noted  that available studies do not produce luminous giants with the cool effective temperature of DY Per and DY Per-variables (see above); predicted luminous giants appear at effective temperatures close to the minimum found for RCB-variables (T$_{\rm eff} \sim 4000$ K) and not at the roughly 3000 K temperature of DY Per.  A likely key ingredient in the modeing of the  white dwarf merger event is the rate at which the He white dwarf is merged with the C-O white dwarf.  Zhang et al. (2014) explore a model in which the merger and the associated nucleosynthesis  is complete in  only about  500 years yet the luminous RCB star  takes almost 1000,000 yrs to evolve through the EHe stage.  Munson, Chatzopoulos \& Denissenkov (2022, Fig. 14)  highlight that this plausible merger process results in the RCB beginning life rich in $^{14}$C  not the expected $^{14}$N. In the immediate prior stage, almost all $^{14}$N experienced neutron capture $^{14}$N$(n,p)^{14}$C. The RCB is now set to experience radioactive dating as $^{14}$C decays with its half-life of 5730 yr and $^{14}$N builds to its final abundance by at least 1 dex over a period of several $^{14}$C half-lives. Although the 2-0 $^{14}$C$^{16}$O band is shifted red and amongst bands of all other CO isotopologues (Pavlenko, Yurchenko \& Tennyson 2020), a prediction that $^{14}$N existed slightly earlier as $^{14}$C will surely intrigue prospectors. An interesting time is ahead!

\begin{acknowledgments}
We thank Kyle Kaplan and Jacob McLane for help observing with IGRINS at the W.J.
McDonald Observatory. We appreciate the support provided by Greg Mace of the
IGRINS team in Austin. The Immersion Grating INfrared Spectrometer (IGRINS) was
developed under a collaboration between the University of Texas at Austin and
the Korea Astronomy and Space Science Institute (KASI) with  financial support
from the W.J. McDonald Observatory, the US National Science Foundation under
grant AST-1229522 to the University of Texas at Austin, and of the Korean GMT
Project of KASI. DAGH acknowledges support from the ACIISI, Gobierno de
Canarias, and the European Regional Development Fund (ERDF) under a grant with
reference PROID2020010051 as well as the State Research Agency (AEI) of the
Spanish Ministry of Science and Innovation (MICINN) under grant
PID2020-115758GB-I00. This article is based upon work from COST Action
NanoSpace, CA21126, supported by COST (European Cooperation in Science and
Technology). NKR thanks the Instituto de Astrof\'{\i}sica de Canarias for
inviting him as a Severo Ochoa visitor during 2016 April-May when part of this
work was done. NKR would also like to thank An\'{\i}bal Garc\'{\i}a-Hern\'andez
and Arturo Manchado for their kind hospitality during his visit to Tenerife.
\end{acknowledgments}




\end{document}